\documentclass[draftclsnofoot,12pt,onecolumn,doublespace]{IEEEtran}
\usepackage{graphicx}
\usepackage{epsfig}
\usepackage[cmex10]{amsmath}
\usepackage{array}
\usepackage{fixltx2e}
\usepackage{cite}

\usepackage{amsfonts}
\usepackage{amssymb}

\begin{document}
\title{Optimal Shape-Gain Quantization for Multiuser MIMO Systems with Linear Precoding}

\author{Muhammad N. Islam,�\IEEEmembership{ Student Member, �IEEE,
} Raviraj Adve,�\IEEEmembership{ Senior Member, IEEE},
 Behrouz Khoshnevis,~\IEEEmembership{Member, IEEE}
\thanks{Muhammad N. Islam is with WINLAB, Rutgers - State
University of New Jersey, New Jersey, USA. email: mnislam@winlab.rutgers.edu.
Raviraj Adve and Behrouz Khoshnevis are with Electrical \& Computer Engineering, University
of Toronto, Toronto, Canada. email: rsadve@comm.utoronto.ca, bkhoshnevis@comm.utoronto.ca}}

\maketitle


\begin{abstract}
This paper studies the optimal bit allocation for shape-gain vector
quantization of wireless channels in multiuser (MU) multiple-input
multiple-output (MIMO) downlink systems based on linear precoding. Our
design minimizes the mean squared-error between the original and
quantized channels through optimal bit allocation across shape
(direction) and gain (magnitude) for a fixed feedback overhead per
user. This is shown to significantly reduce the quantization error,
which in turn, decreases the MU interference. This paper makes three
main contributions: first, we focus on channel gain quantization and
derive the quantization distortion, based on a Euclidean distance
measure, corresponding to singular values of a MIMO channel. Second, we
show that the Euclidean distance-based distortion of a unit norm
complex channel, due to shape quantization, is proportional to
$2^{-\frac{2 B_s}{2 M - 1}}$, where, $B_s$ is the number of shape
quantization bits and $M$ is the number of transmit antennas. Finally,
we show that for channels in complex space and allowing for a large
feedback overhead, the number of direction quantization bits should be
approximately $\left(2M-1\right)$ times the number of channel magnitude
quantization bits.
\end{abstract}




\section{Introduction}

The availability of channel state information (CSI) at the transmitter
significantly improves the performance of multiuser (MU) multiple-input
multiple-output (MIMO) systems~\cite{Boccardi:a,Love:c,Jindal:a}.
Specifically, CSI is essential for effective communications in the MU
downlink. In frequency division duplex systems, in order to provide the
base station (BS) with CSI, the receivers need to quantize the CSI and
feed the quantized information back to the BS. Clearly this feedback is
an overhead to the system and, therefore, must be limited to an
acceptable level.

This paper focuses on limited-feedback MU MIMO systems, where a single
BS communicates with multiple receivers and each user can potentially
receive multiple data streams. Specifically, we restrict our analysis
to systems using linear precoding~\cite{Khachan}. Our main goal is to
present an efficient quantization scheme for these systems. For this
purpose, we use sum mean squared error across all data streams as the
design objective and focus on quantization issues by assuming perfect
channel estimation at the receiver (user) side and a noiseless,
delay-free, feedback link to the BS.

This work is mainly motivated by the fact that the performance of
limited-feedback MU-MIMO systems is very sensitive to the quality of
the CSI available at the BS. Without accurate CSI, the quantization
error common performance measures saturate in the high
signal-to-noise ratio (SNR) regime because the BS cannot completely
pre-cancel the multi-user interference~\cite{Jindal:a,Love:a,Nazmul:b}.
It is therefore essential to design a limited feedback system such that
the CSI quantization error is minimized. The optimal design of channel
quantization in MU-MIMO systems is, therefore, the core objective of
this paper.

Most of the works in limited feedback literature focus on either channel 
magnitude quantization~\cite{skoglund,Aazhang:b,Sabharwal} or 
direction quantization~\cite{Boccardi:a,Jindal:a,Wornell,Erkip} but not both.
However, in MIMO systems, optimizing the precoder at the transmitter depends on
both the channel magnitude (also known as the channel gain) and the
phase of individual channel entries (the channel direction or shape).
The authors of~\cite{Goldsmith} specifically showed that one
needs both channel gain and quantized direction information to achieve
multi-user diversity gain. However, the gain information 
was assumed perfect in~\cite{Goldsmith}.
In general, joint vector quantization (VQ) of channel
magnitude and direction is a very complex task~\cite{Lau}. To reduce
the design complexity, the authors of~\cite{Hamkins} investigate
independent quantization of the channel gain and shape and develop
optimal bit allocation across gain and shape of real channel vectors
using spherical codes. The authors of~\cite{Behrouz:a,Behrouz:b} also
use such a product codebook and solve for the optimal bit allocation to
minimize the average transmit power with quality-of-service
constraints. Such a structure has several practical advantages and
provides an analytically tractable framework to optimize the limited
feedback~\cite{Gray,Hamkins,Behrouz:a}. We adopt a similar approach
where channel gain and shape are independently quantized.

The optimal quantizer depends on the transmission scheme and
performance measure used. Due to its simplicity and efficiency, we
adopt an eigen-based combining (EBC) approach to precode the
data~\cite{Boccardi:a,Love:c}. We study quantization of CSI to minimize
the sum mean squared error (SMSE) over all data streams received, a
popular measure in the MU MIMO downlink~\cite{Khachan,Schubert:a}. 
As shown in our earlier work~\cite{Nazmul:c}, there is a one-to-one relationship between the
SMSE objective and the variance of the quantization error. The current
work, therefore, focuses on optimizing the bit allocation across shape
and gain given a budget for feedback overhead per user. Once this bit
allocation is optimized, one can use our earlier work
in~\cite{Nazmul:c} for designing the limited-feedback system. 
To the best of our knowledge, the problem of optimal bit allocation in 
shape-gain vector quantization to minimize the SMSE 
of a multiuser MIMO system has not been investigated before.

This paper makes two key contributions:
\begin{enumerate}
\item We show that the quantization distortion of a uniformly
    distributed unit-norm vector in $\mathbb{C}^M$ is upper bounded
    by: $ K_s \times 2^{-\frac{2 B_s}{2M - 1}}$, where $M$ is the
    total number of transmit antennas, $B_s$ is the number of shape
    quantization bits and $K_s$ is a constant that does not depend
    on $B_s$.

\item We also show that, for channels in complex space, the optimal
    number of channel direction quantization bits should be
    approximately $\left(2M-1\right)$ times the optimal number of
    channel magnitude quantization bits.

\end{enumerate}

Numerical simulations suggest that the proposed bit allocation 
laws provide a substantial improvement over full shape quantization or full gain quantization 
in the SMSE and bit error rate (BER) performance of a multiuser MIMO system. 

\emph{Notation}: Lower case letters denote scalar values while lower
case bold face letters represent column vectors. Upper case boldface
letters denote matrices. The superscripts $ (\cdot)^T $ and $ (\cdot)^H
$ denote the transpose and conjugate transpose operators respectively.
$\mathrm{tr}[\cdot$] denotes the trace operator. $ \mathbf{I} $ is
reserved for the identity matrix whereas $ \mathbf{1} $ represents the
column vector with all one entries. $ diag(a_1,\,\cdots,\,a_n) $
denotes the diagonal matrix with diagonal elements
$a_1,\,\cdots,\,a_n$; whereas $
diag(\mathbf{A}_1,\,\cdots,\,\mathbf{A}_n)$ represents the block
diagonal matrix with the matrices $\mathbf{A}_1,\,\cdots,\mathbf{A}_n$
on its main diagonal. $ ||\cdot||_1 $ denotes the $ L_1 $ norm of the
vector. $ E(\cdot) $ and $S(\cdot)$ denote statistical expectation and
surface area respectively.

The remainder of the paper is organized as follows.
Section~\ref{sec:Model} outlines the limited feedback MIMO system model
and the corresponding shape-gain product VQ structure.
Section~\ref{sec:Analysis} derives the distortion measures and provides
the optimal bit allocation solution; in general, proofs are deferred to
appendices. This section also presents the linear precoding algorithm
that incorporates the optimal bit allocation policy.
Section~\ref{sec:Simulations} presents results of numerical simulations
illustrating the theory developed. The paper wraps up with some
conclusions in Section~\ref{sec:conc}.

\section{System Model}
\label{sec:Model}

We begin by developing the system model for linearly-precoded MU-MIMO
system followed by the model for CSI feedback and the product
shape-gain quantization structure.

\subsection{MU MIMO System Model}
Consider a single base station equipped with $ M $ transmit antennas
communicating with $K$ independent users. User $k$ has $N_k$ antennas
and receives $L_k$ data streams. All data streams are independent of
each other. Let $L = \sum_k L_k$, $N = \sum_k N_k$. To ensure
resolvability, we require $L \leq M $ and $ L_k \leq N_k $.

\begin{figure}[t]
 \begin{center}
 \epsfig{figure=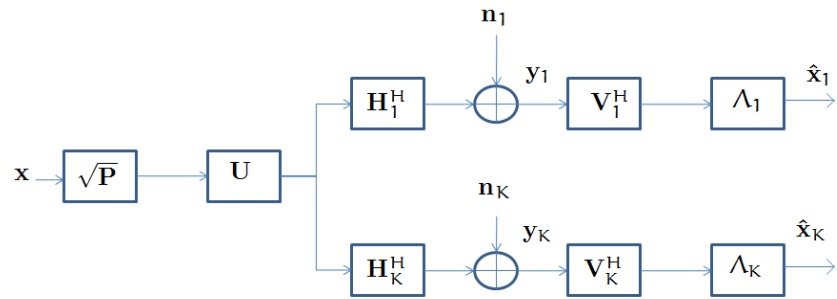,width=130mm}
 \end{center}
  \caption{MU MIMO system model in the downlink}  \label{fig:system_model}
\end{figure}

Let $\mathbf{U} \in \mathcal{C}^{M \times L}$ denote the global
precoder, the columns of which are unit-norm. Similarly, let
$\mathbf{P} \in \mathcal{R}^{L \times L}$ denote the diagonal power
matrix whose entries are the powers allocated to individual streams.
Let, $P_{\max}$ be the total available power; we require
$\mathrm{tr}[\mathbf{P}] \leq P_{\max}$. The data vector $ \mathbf{x} =
\left [x_1, ....,x_L \right ]^T
 = \left[ \mathbf{x}_1^T, \mathbf{x}_2^T, \dots,
\mathbf{x}_K^T\right]^T$, includes all $L$ data streams to the $K$
users. The $N_k \times M$ block fading channel, $\mathbf{H}_k^H$,
between the BS and user $k$ is assumed to be flat. The global channel
matrix is $\mathbf{H}^H $, with $ \mathbf{H} = \left [
\mathbf{H}_1,...,\mathbf{H}_k \right] $. The elements of channel
entries are assumed to be zero mean complex Gaussian random variables
with unit variance. User $ k $ receives
\begin{equation}
\mathbf{y}^{DL}_k = \mathbf{H}^H_k \mathbf{U} \mathbf{\sqrt{P}} \mathbf{x} + \mathbf{n}_k,
\end{equation} %
where $ \mathbf{n}_k $ represents the zero mean additive white Gaussian
noise at the receiver. User $k$, in order to estimate its own
transmitted symbols from $\mathbf{y}^{DL}_k $, forms
\begin{equation}
\hat{\mathbf{x}}_k = \mathbf{\Lambda}_k \mathbf{V}^H_k \mathbf{y}^{DL}_k,
\end{equation}
where $ \mathbf{V}_k \mathbf{\Lambda}_k$ is the $ N_k \times L_k$
decoder matrix for user $ k $. The columns of $\mathbf{V}_k \in
\mathcal{C}^{N_k \times L_k}$ are unit norm while $\mathbf{\Lambda}_k =
diag \left(\lambda_{k_1},\lambda_{k_2},\cdots,\lambda_{k_{L_k}}\right)
\in \mathcal{R}^{L_k \times L_k}$ contains the gain variables that
normalize the received data. Although the gain variables at the
receiver side do not affect the signal-to-interference-plus-noise ratio
(SINR), they play an important role in the error performance of
transmissions that include amplitude modulation, e.g., quadrature
amplitude modulated systems. Figure~\ref{fig:system_model} illustrates
the proposed downlink system.

Let $ \mathbf{V} \mathbf{\Lambda} $ be the $N \times L$ block diagonal
global decoder matrix, $ \mathbf{V} = diag \left ( \mathbf{V}_1, ...,
\mathbf{V}_K \right) \in \mathcal{C}^{N \times L}$ and $
\mathbf{\Lambda} = diag \left ( \mathbf{\Lambda}_1, ...,
\mathbf{\Lambda}_K \right) \in \mathcal{R}^{L \times L}$. Overall,
\begin{eqnarray}
\hat{\mathbf{x}} & = & \mathbf{\Lambda} \mathbf{V}^H\mathbf{H}^H\mathbf{U}
                        \sqrt{\mathbf{P}}\mathbf{x} + \mathbf{V}^H\mathbf{n} \nonumber\\
& = & \mathbf{F}^H \mathbf{U} \sqrt{\mathbf{P}} \mathbf{x} + \mathbf{V}^H \mathbf{n},
\end{eqnarray}
where, $ \mathbf{n} = \left[ \mathbf{n}_1^T, \mathbf{n}_2^T, \dots,
\mathbf{n}_K^T\right]^T$. For the ease of representation, we define the
$M\times L$ matrix $\mathbf{F} = \mathbf{HV\Lambda} $ with $ \mathbf{F}
= \left [\mathbf{f}_1, \dots , \mathbf{f}_L \right]$. The vectors $
\mathbf{f}_1 , \dots, \mathbf{f}_L $ are the effective $ M \times 1 $
vector downlink channels of the individual data streams.

The MSE of the $i^\mathrm{th}$ data stream of the $k^\mathrm{th}$ user
is given by\footnote{Note that we, interchangeably, index streams as
being the $\ell^\mathrm{th}$ of $L$ streams overall or the
$i^\mathrm{th}$ stream of the $k^\mathrm{th}$ user. Any one-to-one
mapping between the two notations is acceptable.},
\begin{equation}
    e^{DL}_{k,i} = E \left [ \left ( \widehat{x}_{k,i} - x_{k,i} \right)
                \left ( \widehat{x}_{k,i} - x_{k,i} \right)^H \right ]. \label{eqn:downlink_mse}
\end{equation}
The min-SMSE optimization problem is:
\begin{equation}
 \min_{\mathbf{p},\mathbf{U}, \mathbf{V}, \mathbf{\Lambda} }
            \sum^K_{k=1} \sum^{L_k}_{i=1} e^{DL}_{k,i}; \hspace*{0.1in} \mathrm{subject\ to}
    \hspace*{0.1in} \mathrm{tr}[\mathbf{P}] \leq P_{\max},
                                                ||\mathbf{u}_\ell||=||\mathbf{v}_\ell||=1,
     \label{objective}
\end{equation}
To solve this problem, it is computationally efficient to use a virtual
dual uplink~\cite{Khachan}. In this uplink the transmit powers are $
\mathbf{Q} = diag \left[q_1, .., q_L \right]^T$ for the $L$ data
streams, while the matrices $ \mathbf{U} $ and $ \mathbf{V} $ remain
the same as before.

\subsection{Feedback Model:}

As mentioned earlier, we use an eigen-mode
strategy~\cite{Boccardi:a,Love:c}. According to this strategy, the
$k^\mathrm{th}$ user estimates its own channel $\mathbf{H}_k$ and uses
a set of dominant singular values and singular vectors of
$\mathbf{H}_k$ as $\mathbf{\Lambda}_k$ and $\mathbf{V}_k$ respectively.

Since the user is aware of $\mathbf{H}_k$, $\mathbf{V}_k$ and
$\mathbf{\Lambda}_k$, it can form the product matrix $\mathbf{F}_k =
\mathbf{H}_k \mathbf{V}_k \mathbf{\Lambda}_k$, whose columns act as the
effective vector downlink channels for the data streams. Each user
quantizes its effective vector downlink channel based on an Euclidean
distance measure and feeds back the quantized channel to the BS.
Details of the CSI quantization policy will be described in the next
section. To model the effect of quantization, we consider the following
relation between the original and the quantized variables,
\begin{equation}
\mathbf{f}_{k,i} = \mathbf{\widehat{f}}_{k,i} + \mathbf{\widetilde{f}}_{k,i}
                                         \hspace*{0.1in} \mathrm{or}
\hspace*{0.1in} \mathbf{F} = \mathbf{\widehat{F}} + \mathbf{\widetilde{F}}.
\label{channel_model}
\end{equation} %
Here, $\mathbf{f}_{k, i}$ denotes the effective vector downlink channel
of the $i^\mathrm{th}$ stream of the $k^\mathrm{th}$ user. $\mathbf{F}
$ comprises $L$ effective channel vectors with the original channel
directions and channel gains. $ \mathbf{\hat{F}} $ denotes the $L$
quantized feedback vectors. The matrix $\mathbf{\tilde{F}}$ represents
the quantization error.

The BS assumes that the quantization error matrix $ \mathbf{\tilde{F}}
$ has $ M \times L $ independent identically Gaussian distributed
(i.i.d.)  elements with zero mean and a variance of $\sigma^2_E/M$,
where $\sigma^2_E$ is the quantization error variance associated with
each quantized vector $ \mathbf{\hat{f}}_{k,i} $ and is defined as,
\begin{equation}
\sigma^2_E = E \left[ ||\mathbf{f}_{k,i} - \mathbf{\widehat{f}}_{k,i} ||^2 \right].
\end{equation}

By using the optimal $\mathbf{P}$ and $\mathbf{U}$, the minimum SMSE
takes the following form~\cite{Nazmul:c}:
\begin{equation}
SMSE = L - M + \left (\sigma^2 +\frac{\sigma^2_E}{M} P_{\max} \right)
                \mathrm{tr}\left [ \mathbf{J}^{-1} \right], \label{eq:SMSE}
\end{equation}
where,
\begin{equation}
\mathbf{J} = \mathbf{\hat{F}} \mathbf{Q} \mathbf{\hat{F}}^H + \left(\sigma^2 +
                                \frac{\sigma^2_E}{M} P_{\max} \right) \mathbf{I}_M.
\label{eq:J}
\end{equation}
where $\mathbf{Q}$ is the virtual uplink power allocation matrix.

Equations in~\eqref{eq:SMSE} and~\eqref{eq:J} show that the SMSE is
directly related to the quantization error $\sigma^2_E$. The limited
feedback system design problem can therefore be formulated as
minimization of the quantization error variance subject to a fixed
feedback overhead.

\subsection{Shape-Gain Product Quantization Model}

We intend to find the optimal bit allocation for quantizing the
effective vector downlink channel $\mathbf{f}_{k,i}$. From now on, we
will use $\mathbf{z}$ to represent the effective vector downlink
channel to simplify the notation. According to the eigen-based receiver
structure assumed in this work, $\mathbf{z}$ represents the product of
a singular value of the channel matrix and its corresponding singular
vector.

\begin{figure}[t]
 \begin{center}
 \epsfig{figure=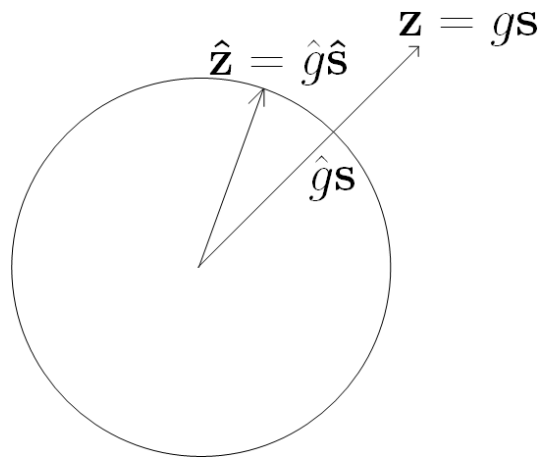,width=70mm}
 \end{center}
  \caption{Gain-shape product quantization}  \label{fig:product_quantization}
\end{figure}

Let $\mathbf{\hat{z}}$ be the quantized effective vector downlink
channel and let $\mathcal{C} = \left\{
{\mathbf{c}_1,\mathbf{c}_2,\cdots,\mathbf{c}_{N_{tot}}} \right\}$
denote the codebook of quantized channels. Here, $N_{tot} = 2^B$ are
the total number of quantization levels using a total of $B$ bits. This
codebook is simplified to a product codebook.
Fig.~\ref{fig:product_quantization} illustrates the product codebook
operation based on independent quantization of gain and shape. Let,
$\mathbf{z} = g \mathbf{s}$ where, $g = ||\mathbf{z}||_2$ and the
unit-norm $\mathbf{s} = \mathbf{z}/||\mathbf{z}||_2$ denote the gain
and shape of the channel respectively. The BS is provided with the
quantized information $\mathbf{\hat{z}} = \hat{g} \hat{\mathbf{s}}$,
where $\hat{g}$ and $\hat{\mathbf{s}}$ denote the quantized gain and
shape respectively.

Let $B_g$ and $B_s$ denote the number of bits allocated to gain and
shape quantization and define $N_g = 2^{B_g}$ and $N_s = 2^{B_s}$.
Further, let $\mathcal{C}_g$ and $\mathcal{C}_s$ represent the gain and
shape codebook respectively:
\begin{eqnarray}
\mathcal{C}_g & = & \left[c_{g_1},c_{g_2},\cdots,c_{g_{N_g}} \right]
                                        \label{eq:gain_codebook}  \\
\mathcal{C}_s & = & \left[ {\mathbf{c}_{s_1},\mathbf{c}_{s_2},\cdots,
                        \mathbf{c}_{s_{N_s}}} \right]. \label{eq:shape_codebook}
\end{eqnarray}
The product codebook can therefore be represented as,
\begin{equation}
\mathcal{C} = \mathcal{C}_g \times \mathcal{C}_s.   \label{eq:main_codebook2}
\end{equation}
The quantized gain and shape variables are computed as:
\begin{eqnarray}
\hat{g} & = & \arg \min_{c_g \in \mathcal{C}_g} (g-c_g)^2
                        \label{eq:generic_gain_quantization}  \\
\mathbf{\hat{s}} & = & \arg \min_{\mathbf{c}_s \in \mathcal{C}_s} ||\mathbf{s}-\mathbf{c}_s||^2.
                                            \label{eq:generic_gain_quantization1}
\end{eqnarray}

The Lloyd-Max algorithm is the optimal solution to find the codebook
for the gain of the channel vector with the MSE objective~\cite{Lloyd}.
We use the $K$-means approach, as described in~\cite{Kmeans}, for
numerical implementation of the Lloyd-Max algorithm. The optimal
codebook of unit norm vectors with a Euclidean measure is not yet
known. Therefore, we adopt random VQ to find the shape codebook. With
this approach, the unit norm quantized shape vectors are randomly and
independently distributed on the complex unit hyper-sphere in
$\mathbb{C}^{M}$.

The remaining question is, given $B$, what is the optimal choice of
$B_s$ and $B_g$?

\section{Distortion analysis and optimal bit allocation solution}
\label{sec:Analysis}

\subsection{Design Objective}

Our main problem is to optimize the shape-gain bit allocation as formulated below,
\begin{align}
[B_s^*, B_g^*] = \arg \min_{Bs,Bg} & E \left [||\mathbf{z}-\hat{g} \mathbf{\hat{s}}||^2
                                                \right]   \label{eq:bit_alloc_problem2}  \\
subject \, to: \, \, & B_s + B_g = B \, , \, B_s \geq 0 \, , \, B_g \geq 0 \, , \,
        \hat{g} \in \mathcal{C}_g \, , \, \mathbf{\hat{s}} \in \mathcal{C}_s. \nonumber
\end{align}
Hamkins et al.~\cite{Hamkins} have shown that, for high resolution quantization (large
$B_s$ and $B_g$), the distortion measure takes the following
form~\cite{Hamkins}:
\begin{align}
E \left [||\mathbf{z}-\hat{g} \mathbf{\hat{s}}||^2 \right] &
    \approx E \left [(g-\hat{g})^2 \right] + E \left[ g^2 \right]
    E \left [||\mathbf{s} - \mathbf{\hat{s}}||^2 \right]  \allowdisplaybreaks[1] \\
& \approx D_g + E \left[ g^2 \right] D_s \allowdisplaybreaks[1],  \label{eq:overall_dist}
\end{align}
where, $E \left[ g^2 \right]$ denotes the variance of the gain and $D_g
=  E \left [(g-\hat{g})^2 \right]$ is the gain quantization distortion.
On the other hand, $D_s = E \left [||\mathbf{s} - \mathbf{\hat{s}}||^2
\right]$ represents the distortion due to unit-norm shape quantization.
Since $D_g$ and $D_s$ are independent of each other
in~\eqref{eq:overall_dist}, the optimal bit allocation problem can be
solved using the following three steps:
\begin{enumerate}
\item Find $D_g$, gain distortion, for a given $B_g$.

\item Find $D_s$, shape distortion, for a given $B_s$.

 \item Provide optimal bit allocation to minimize the overall
     distortion, i.e., $ E \left[ g^2 \right] D_s + D_g $.
\end{enumerate}

\subsection{Distortion due to Gain Quantization}

The distortion due to quantizing the gain is given by
\begin{align}
D_g = E \left [(g-\hat{g})^2 \right]  = \int^{\infty}_{0} \left (r - \hat{g}(r) \right)^2 f_g (r) dr.
\label{eq:distortion2} \allowdisplaybreaks[1]
\end{align}
Here, $\hat{g}(r)$ is the quantized value of $r$ and $f_g (r)$ is the
probability density function (pdf) of the gain. Using Bennett's
integral (~\cite{Gray}, page-186), the distortion
in~\eqref{eq:distortion2} takes the form,
\begin{equation}
D_g = \frac{1}{12 N_g^2} ||f_g (r)||_{\frac{1}{3}},  \label{eq:distortion4}
\end{equation}
where, $N_g = 2^{B_g}$ and
\begin{equation}
||f_g (r)||_{\frac{1}{3}} = \left( \int^{\infty}_0 |f_g(r)|^{\frac{1}{3}} dr \right)^3.
\end{equation}
\emph{Lemma 1:} For Rayleigh fading and based on the pdf of the
dominant eigenvalues of Wishart matrix and Jacobian transformation
in~\cite{Karasawa} and~\cite{Strang}, we have,
\begin{equation}
||f_g (r)||_{\frac{1}{3}} = \frac{3 \times 3^{L(e)} \beta}{4 (L(e) - 1)!}
\Gamma^3 \left(\frac{L(e) + 1}{3} \right),
\label{eq:distortion5}
\end{equation}
where, $L(e) = (M-e) (N_k-e)$, $M$ represents the total number of transmit antennas at the BS,
 $N_k$ denotes the number of receiver antennas of  the $k^\mathrm{th}$
user. $e$ denotes the index of the ordered eigenvalues where $0$
represents the most dominant one, $1$ denotes the 2nd most dominant one
and so on. Finally, $\beta = \tilde{\lambda_e}/L(e)$ where
$\tilde{\lambda_e}$ is the mean of the $\mathrm{e}^{th}$ eigenvalue.

\emph{Proof:} See Appendix A. $\blacksquare$

Using \eqref{eq:distortion4} and \eqref{eq:distortion5}, the gain
distortion at high resolution can be expressed as,
\begin{align}
D_g & = \frac{1}{12 N_g^2}  ||f_g (r)||_{\frac{1}{3}} \allowdisplaybreaks[1]  \\
& = \frac{1}{16 N_g^2} \frac{3^{L(e)} \beta}{(L(e) - 1)!} \Gamma^3
                        \left(\frac{L(e) + 1}{3} \right) \allowdisplaybreaks[1]  \\
& = K_g 2^{-2 B_g},   \label{eq:distortion_MIMO} \allowdisplaybreaks[1]
\end{align}
where,$ K_g = \frac{1}{16} \frac{3^{L(e)} \beta}{(L(e) - 1)!} \Gamma^3
\left(\frac{L(e) + 1}{3} \right) $ is a constant with respect to $ B_g
$. Equation~\eqref{eq:distortion_MIMO} suggests that the gain
distortion due to quantization is proportional to $2^{-2 B_g}$.

\begin{figure}[t]
 \begin{center}
 \epsfig{figure=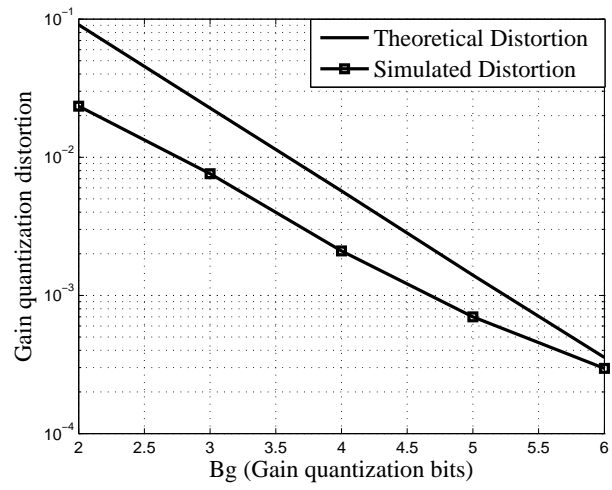,width=90mm}
 \end{center}
  \caption{Quantization distortion of the dominant singular value of 2x2 MIMO channel}
  \label{fig:gain_quantization}
\end{figure}

Figure~\ref{fig:gain_quantization} shows the distortion due to gain
quantization of the dominant singular value of a $2 \times 2$ MIMO
channel. As the figure verifies, the analytical expression converges to
 the simulation result as $B_g$ increases.

\subsection{Shape Quantization Distortion}

This section focuses on the shape quantization distortion of a
unit-norm vector in $\mathbb{C}^M$, in terms of the Euclidean distance.
The Euclidean distance of two points in a $\mathbb{C}^M$ plane has a
one-to-one relation with the distance of two points in a
$\mathbb{R}^{2M}$ plane. Therefore, we can focus on quantization of
unit-norm vectors in $\mathbb{R}^{2M}$ instead of $\mathbb{C}^M$.

\begin{figure}[t]
 \begin{center}
 \epsfig{figure=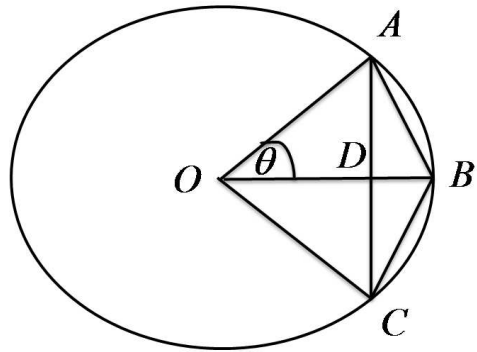,width=80mm}
 \end{center}
  \caption{Shape quantization block diagram}  \label{fig:shape_quantization}
\end{figure}

Figure~\ref{fig:shape_quantization} shows a two dimensional view of the problem where
$OB = \mathbf{s}$, $OA = \mathbf{\hat{s}}$. Here, $||\mathbf{s}||_2 = ||\mathbf{\hat{s}}||_2 = 1$.
The Euclidean distance between $\mathbf{s}$ and $\mathbf{\hat{s}}$ is defined by,
$ d = ||\mathbf{s}-\mathbf{\hat{s}}||_2$.  Define $\mathcal{U}_{2M}$ as the unit
hypersphere in $\mathbb{R}^{2M}$. The surface area of $\mathcal{U}_{2M}$
is given by~\cite{Behrouz:a}
\begin{equation}
S \left(\mathcal{U}_{2M}\right) = 2M C_{2M},   \label{eq:area_hypersphere}
\end{equation}
where,
\begin{equation}
C_{2M} = \frac{\pi^M}{\Gamma(M+1)}.     \label{eq:C_2M}
\end{equation}
Define the spherical cap $\mathcal{D}$, i.e., the region $ABC$ around
$\mathbf{s}$ in Fig.~\ref{fig:shape_quantization}, as:
\begin{equation}
\mathcal{D} = \left(\mathbf{\hat{s}} \in \mathcal{U}_{2M} | ||\mathbf{s}-\mathbf{\hat{s}}||_2
\leq d \right), \label{eq:def_cap}
\end{equation}
and let $\angle AOB = \theta$ be the angular distance between
$\mathbf{s}$ and $\mathbf{\hat{s}}$. Since $ ||OA||_2 =
||\mathbf{\hat{s}}||_2 = 1$, we have $AD = \sin(\theta)$ and $OD =
\cos(\theta)$. Also, since $||OB||_2 = ||\mathbf{s}||_2 = 1$, we have
$BD = 1 - \cos(\theta)$. Therefore,
\begin{equation}
 AB^2 = AD^2 + BD^2 = \sin^2(\theta) + \left(1 - \cos(\theta)\right)^2 =
                            2 - 2 \cos(\theta).  \label{eq:rel_angle_euclidean1}
\end{equation}
Here, if we define $b = d^2$, we will have:
\begin{align}
\theta & = \cos^{-1} \left(1-0.5 b\right).  \label{eq:rel_angle_euclidean3} \allowdisplaybreaks[1]
\end{align}
The surface area of $\mathcal{D}$ is given by~\cite{Behrouz:a},
\begin{equation}
S \left(\mathcal{D} \right) = (2M-1) C_{2M-1} \int^{\theta}_0 \sin^{2M-2} \phi \, d\phi.
                \label{eq:area_cap}
\end{equation}
Now, if we assume a small spherical cap of radius $d$ centered on
$\mathbf{s}$, the quantized vector can lie anywhere on this cap. Hence,
\begin{equation}
Pr [||\mathbf{s}-\mathbf{\hat{s}}||^2 \leq b] = \frac{S (\mathcal{D})}{S (\mathcal{U}_{2M})}.
\label{eq:rel_probability_area}
\end{equation}
Using \eqref{eq:area_hypersphere}, \eqref{eq:C_2M},
\eqref{eq:rel_angle_euclidean3} and \eqref{eq:area_cap} in
\eqref{eq:rel_probability_area} we get
\begin{equation}
Pr [||\mathbf{s}-\mathbf{\hat{s}}||^2 \leq b] =  \frac{(2M-1) C_{2M-1}
\int^{\cos^{-1} \left(1-0.5 b\right)}_0 \sin^{2M-2} \phi \, d\phi}{2M C_{2M}}.
\label{eq:less_than_prob2}
\end{equation}
Since all the quantized vectors are randomly chosen, the probabilities
that the square of the Euclidean distance between any vector in the
codebook and the corresponding channel is higher than $b$, are
independent of each other. Therefore,
\begin{equation}
Pr [ \min_{i \in [1,N_s]} ||\mathbf{s}-\mathbf{\hat{s}}_i||^2 \geq b  ]  =
\left(1 - \frac{(2M-1) C_{2M-1} \int^{\cos^{-1} \left(1-0.5 b\right)}_0
\sin^{2M-2} \phi d\phi}{2M C_{2M}} \right)^N.    \label{eq:CCDF1_old}
\end{equation}
Hence, expected value of the distortion error due to shape quantization
can be calculated as follows:
\begin{equation}
E(b)  =  \int_0^4 Pr [ \min_{i \in N} ||\mathbf{s}-\mathbf{\hat{s}}_i||^2 \geq b  ] db.
\label{eq:expect}
\end{equation}
The limits of integration in \eqref{eq:expect} follow from the fact
that the square of the Euclidean distance between two points on a unit
radius sphere has a range of $0$ to $4$.

\emph{Lemma 2}:
\begin{equation}
E (b) < K_s 2^{\frac{-2 B_s}{2M-1}},   \label{eq:lemma_expectation}
\end{equation}
where,
\begin{equation}
K_s = \left(\frac{\pi^{\frac{2M-1}{2}} \Gamma(M)}{2 \pi^M \Gamma
                    \left(\frac{2M-1}{2} + 1 \right)} \right)^{\frac{-2}{2M-1}}.
\label{eq:K_s}
\end{equation}
is a constant.

Proof: See Appendix B. $\blacksquare$

Figure~\ref{fig:shape_distortion_slope} shows that the upper bound of
the shape distortion in \eqref{eq:lemma_expectation} has a fixed gap
with respect to the simulation result. Therefore, we can safely
approximate the shape distortion with the analytical expression in
\eqref{eq:shape_distortion_final2}. Thus,
\begin{equation}
D_s = E \left(||\mathbf{s}-\mathbf{\hat{s}}||^2 \right) \approx
\left(\frac{\pi^{\frac{2M-1}{2}} \Gamma(M)}{2 \pi^M
\Gamma \left(\frac{2M-1}{2} + 1 \right)} \right)^
{\frac{-2}{2M-1}} 2^{\frac{-2 B_s}{2M-1}} \allowdisplaybreaks[1] =  K_s \,
                2^{\frac{-2 B_s}{2M-1}}. \label{eq:shape_distortion_final2}
\end{equation}

\subsection{Optimal Bit Allocation}

Having analyzed the individual terms in~\eqref{eq:overall_dist}, we are
now able to answer the core question of this paper: the allocation of
bits across gain and shape. In~\eqref{eq:overall_dist}, the overall
distortion measure was shown to take the following form,
\begin{equation}
D = E \left [||\mathbf{z}-\hat{g} \mathbf{\hat{s}}||^2 \right]
\approx D_g + E \left[ g^2 \right] D_s.
\label{eq:overall_dist0}
\end{equation}
Using the gain and shape distortion measures
of~\eqref{eq:distortion_MIMO} and~\eqref{eq:shape_distortion_final2},
$D$ can be approximated as,
\begin{align}
D & \approx E \left[ g^2 \right] K_s 2^{-\frac{2 B_s}{2 M - 1}} + K_g 2^{-2 B_g}
                                                            \allowdisplaybreaks[1] \\
& \approx \bar{K_s} 2^{-\frac{2 B_s}{2 M - 1}} + K_g 2^{-2(B - B_s)},
                            \allowdisplaybreaks[1] \label{eq:distortion}
\end{align}
where $\bar{K_s} = K_s E \left[ g^2 \right]$. With these relations in
hand, the optimal shape-gain bit allocation can be formulated as
follows,
\begin{equation}
B_s^* =
\arg \min_{B_s} \bar{K_s} 2^{-\frac{2 B_s}{2 M - 1}} + K_g 2^{-2(B - B_s)}
\end{equation}
\emph{Theorem 1:}  \\
The optimal bit allocation problem has the following solution:
\begin{align}
B_s & = \frac{2M-1}{2M} B + \frac{2M-1}{4M} \log_2 \left(\frac{\bar{K_s}}{K_g (2M - 1)}\right)
\allowdisplaybreaks[1] \label{eq:shape_bits_chap4} \\
B_g = B - B_s & = \frac{1}{2M} B - \frac{2M-1}{4M} \log_2 \left(\frac{\bar{K_s}}{K_g (2M - 1)}\right).
\allowdisplaybreaks[1]  \label{eq:gain_bits_chap4}
\end{align}
Here, $\bar{K_s}$ and $K_g$ are the terms defined in the previous subsections.

\emph{Proof:} See Appendix C. $\blacksquare$

Note that, $\bar{K_s}$ and $K_g$ in \eqref{eq:shape_bits_chap4} and
\eqref{eq:gain_bits_chap4} depend on $M$ but not $B$. Therefore, as $B$
goes to infinity,
\begin{align}
B_s & \approx \frac{2M-1}{2M} B   \label{eq:shape_rate} \allowdisplaybreaks[1]   \\
B_g & \approx \frac{1}{2M} B.        \label{eq:gain_rate} \allowdisplaybreaks[1]
\end{align}
The analytical expressions of \eqref{eq:shape_rate} and
\eqref{eq:gain_rate} can be intuitively explained as follows: The norm
of a $\mathbb{C}^{M}$ vector varies across a one dimensional line.
However, the shape of a $\mathbb{C}^{M}$ vector is uniformly
distributed in the surface of a $(2M-1)$ dimensional hypersphere.
Therefore, given $2M$ number of bits to quantize a $\mathbb{C}^M$
vector, one should expend approximately $1$ and $(2M-1)$ bits to
quantize the gain and shape of the vector respectively. It is worth
noting that, from a different point of view and using a very different
analysis, the work in~\cite{Behrouz:a, Behrouz:b} leads to a similar
expression and explanation. However, this similarity is only for a high
available feedback rate.

Finally, the overall quantization error for a fixed feedback overhead
takes the following form:
\begin{align}
D = & \bar{K_s} 2^{-\frac{2 B_s}{2M-1}} + K_g 2^{-2 B_g} \label{eq:overall_dist1}
\allowdisplaybreaks[1]  \\
= & \bar{K_s} 2^{-\frac{2}{2M-1} \left(\frac{2M-1}{2M} B +
    \frac{2M-1}{4M} \log_2 \left(\frac{\bar{K_s}}{K_g (2M-1)} \right) \right)} +
        K_g 2^{-2 \left(\frac{1}{2M} B - \frac{2M-1}{4M} \log_2 \left(\frac{\bar{K_s}}
                                {K_g (2M-1)} \right) \right)}  \nonumber \allowdisplaybreaks[1] \\
= & 2^{-\frac{B}{M}} \log_2 \left(\frac{\bar{K_s}}{K_g (2M-1)}\right)
    \left(\bar{K_s} 2^{-\frac{1}{2M}} - K_g 2^{-\frac{2M-1}{2M}}\right)
                    \label{eq:overall_dist3} \allowdisplaybreaks[1]  \\
= & D_c 2^{-\frac{B}{M}},  \label{eq:overall_dist4} \allowdisplaybreaks[1]
\end{align}
where, $ D_c = \log_2 \left(\frac{\bar{K_s}}{K_g (2M-1)}\right)
\left(\bar{K_s} 2^{-\frac{1}{2M}} - K_g 2^{-\frac{2M-1}{2M}}\right)$ is
a constant.
%

\subsection{Overall Linear Precoding Algorithm} \label{sec:Algorithm}

In the previous section we derived the optimal allocation of available
bits across gain and shape. Here we use this information to develop the
overall linear precoding algorithm. The material exploits previous work
in~\cite{Nazmul:a,Nazmul:c}. The algorithm steps are:

\begin{enumerate}

\item A gain codebook of $B_g$ bits is generated based on the
    dominant singular values of a random Gaussian matrix and using
    the $K$-means algorithm~\cite{Kmeans}. A shape codebook of
    $2^{B_s}$ random unit-norm vectors, uniformly and independently
    distributed in $\mathbb{C}^M$, is also generated. Both these
    codebooks are generated off-line and the codebooks are shared
    between the BS and the users.

\item The BS sends common pilot symbols so that the receivers can
    estimate $\mathbf{H}_k$.

\item The receivers calculate the dominant singular values
    $\mathbf{\Lambda}_k$ and the corresponding singular vectors
    $\mathbf{V}_k$ and form $\mathbf{F}_k = \mathbf{H}_k
    \mathbf{V}_k \mathbf{\Lambda}_k $.

\item The receivers use the codebooks to quantize the gain and
    shape of the column vectors in $\mathbf{F}_k$ and feedback the
    quantization indices to the BS.

\item The BS calculates the optimum virtual uplink power allocation
    matrix as,
    \[ \mathbf{Q}^{opt} = \min_\mathbf{Q} \left
    (\sigma^2 +\frac{\sigma^2_E P_{\max}}{M}  \right)
    \mathrm{tr}(\mathbf{J}^{-1}) , \, \, \mathrm{such~that} \, \, \mathrm{tr}(\mathbf{Q}) \leq
    P_{\max}.\] Here, $\sigma^2_E = D$ as in
    \eqref{eq:overall_dist4} and $\mathbf{J}$ is calculated
    according to  \eqref{eq:J}.

\item The precoding matrix of the $k^\mathrm{th}$ user is
    calculated as, $\mathbf{U}_k^{mmse} = \mathbf{J}^{-1}
    \hat{\mathbf{F}}_k \sqrt{\mathbf{q}_k} $. Here, $\mathbf{q}_k =
    [q_{k_1},\cdots,q_{k_{L_k}}]$ contains the virtual uplink power
    variables of the $L_k$ streams of the $k^\mathrm{th}$ user.

\item Using a recent result~\cite{Adam:EqualPower}, the downlink
    transmit power variables are determined as, $\mathbf{p} =
    \mathbf{q}$. Here, $\mathbf{q} = \left[q_1,\cdots,q_L \right]$
    is the virtual uplink transmit powers of the $L$ streams.

\end{enumerate}


\section{Numerical Simulations}
\label{sec:Simulations}

\begin{figure} [t]
 \begin{center}
 \epsfig{figure=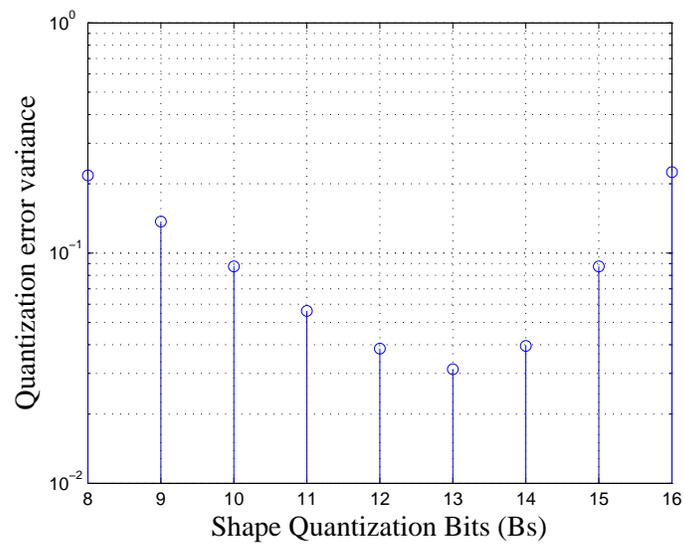,width=100mm}
 \end{center}
  \caption{Effect of bit allocation in the quantization of the product of dominant eigenvalue \& the
  corresponding eigenvector of a 2 x 2 MIMO channel}  \label{fig:MIMO_quantization}
\end{figure}

\begin{figure} [t]
 \begin{center}
 \epsfig{figure=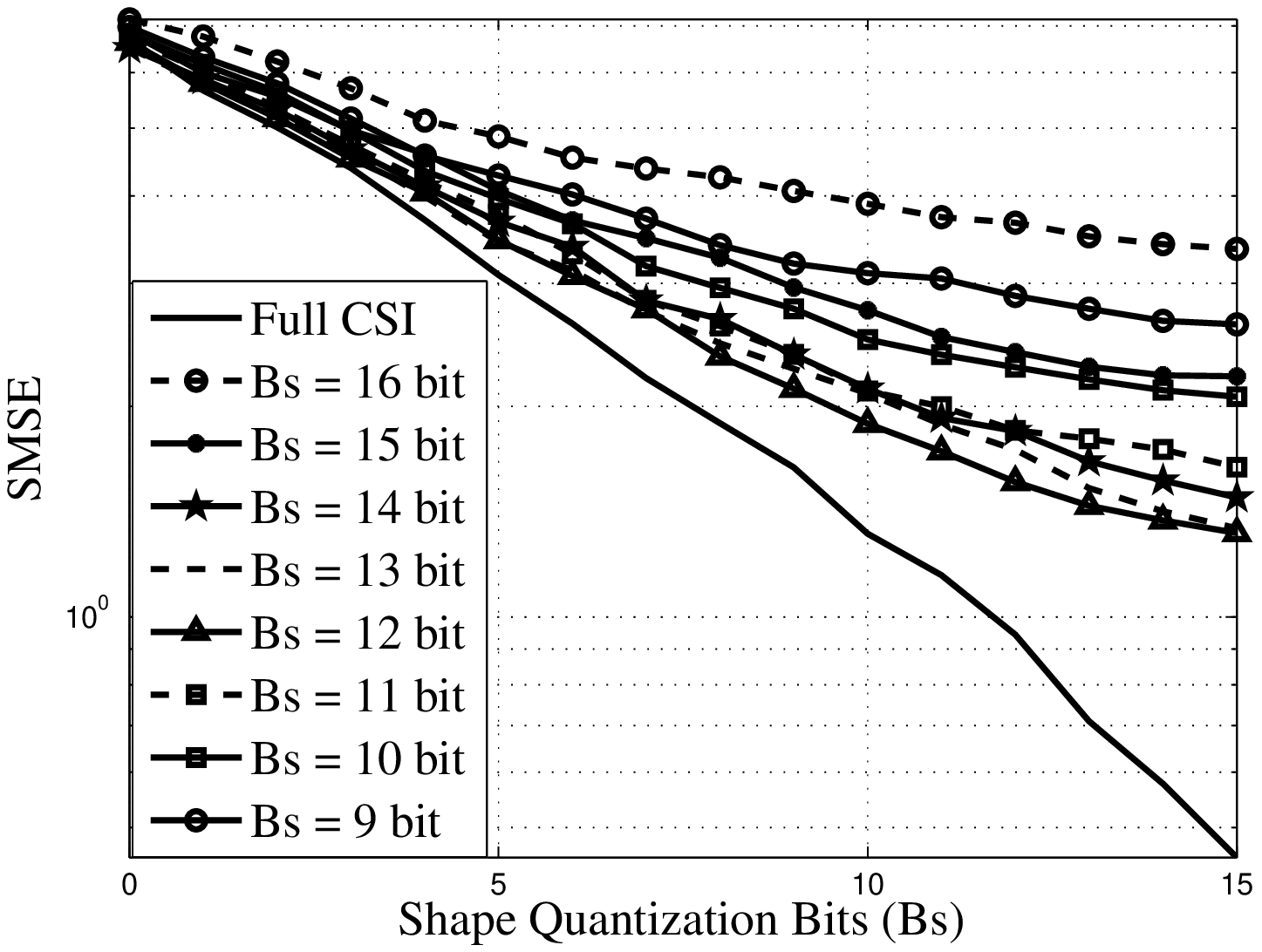,width=100mm}
 \end{center}
  \caption{Effect of bit allocation in the SMSE of 16-QAM system, $M = 2, N = [2~2],
  L = [1~1], B = 16$}  \label{fig:16QAM_SMSE}
\end{figure}

\begin{figure}
 \begin{center}
 \epsfig{figure=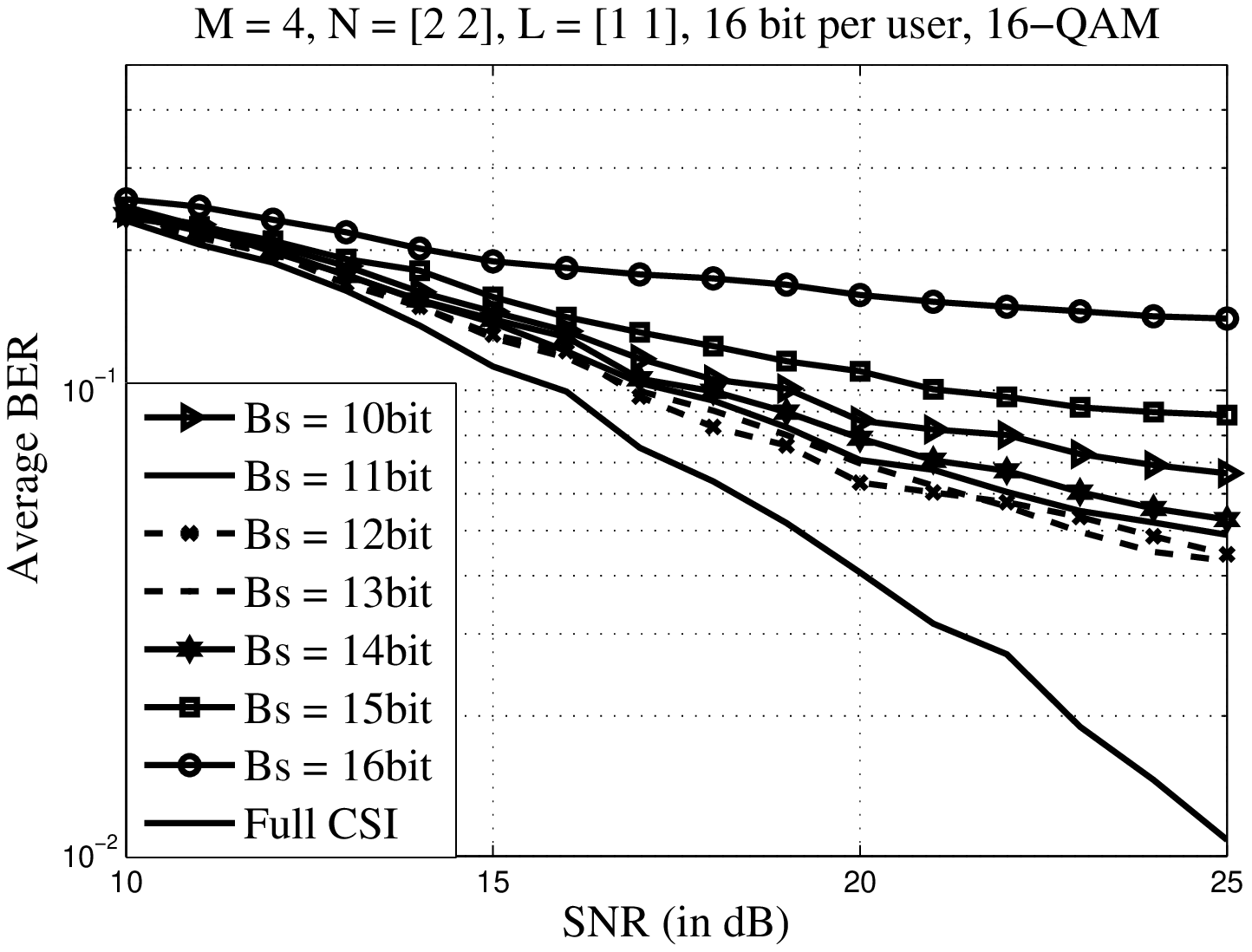,width=100mm}
 \end{center}
  \caption{Effect of bit allocation in the BER of 16-QAM systems, $M = 2, N = [2~2],
  L = [1~1], B = 16$}  \label{fig:16QAM_BER}
\end{figure}

This section provides the results of simulations to study the effect of
shape-gain quantization on the performance of the MU MIMO linear
precoding scheme described. We assume the following scenario in the
simulation setup: the base station has two transmit antennas and serves
two receivers in the downlink. Each receiver has 2 receive antennas and
receives 1 data stream. The feedback overhead per user, $B$, is 16
bits. We show performance curves for different shape quantization bits,
$B_s$, in Figs.~\ref{fig:MIMO_quantization}, \ref{fig:16QAM_SMSE} and
\ref{fig:16QAM_BER}. The corresponding number of gain quantization bits
is given by, $B_g = B - B_s$.

Figure~\ref{fig:MIMO_quantization} illustrates the effect of bit
allocation on the quantization error and suggests that $B_s = 13$ and
$B_g = 3$ are the optimal bit allocations for this scenario. The
analytical results in \eqref{eq:shape_bits_chap4}
and~\eqref{eq:gain_bits_chap4} lead to, $B_g = 2.6$, $B_s = 13.4$ which
matches the numerical result.

Fig.~\ref{fig:16QAM_SMSE} plots the SMSE of the same system with the
transmitter using 16-QAM. The figure shows that $B_s = 12$ leads to the
minimum SMSE. The SMSE performance of $B_s = 13$, i.e., the 
optimal solution obtained from analytical results,
is very close to that of $B_s = 12$ bits.
The minor difference between the simulation and
analytical result stems from the fact that the derived gain distortion
holds only for large number of gain quantization bits. Note that,
$B_s=16$, i.e., quantizing the shape exclusively, leads to much higher
SMSE. Therefore, optimal bit allocation across gain and shape feedback
provides better performance in terms of SMSE.

Fig.~\ref{fig:16QAM_BER} shows that the bit allocation $B_s = 12$ or $B_s = 13$
also lead to better performance in terms of BER. If one uses all the bits
for direction quantization, i.e., $B_s = 16$, the effect of
multiuser interference on the norm of the received signal cannot be
rectified. This leads to the inferior performance of $B_s = 16$.

We did not plot the performance of all possible bit allocations, 
e.g., $B_s = 0$ to $B_s = 8$, so that the figures look clearer. However, 
the performance trend of $B_s = 11$ to $B_s = 9$ bit clearly suggests that 
full gain quantization ($B_s = 0$, $B_g = 16$) will also perform much inferior
to the optimal bit allocation in shape-gain quantization. Thus, optimal 
shape-gain quantization can improve over full gain or full shape quantization and lead
to a lower BER in multiuser MIMO systems. In wireless ethernet~\cite{MAC} systems, where a small
number of bit errors may lead to the whole packet drop~\cite{Park}, optimal bit allocation in shape-gain
quantization can significantly reduce the packet loss rate and save packet re-transmission time.

\section{Conclusion}
\label{sec:conc}

This paper studies the optimal bit allocation across gain and shape
quantization in a MU MIMO downlink system by minimizing the SMSE of the
system for a fixed feedback overhead per user. We show that the
distortion due to gain and shape quantization are proportional to
$2^{-2 B_g}$ and $2^{-\frac{2 B_s}{2M - 1}}$ respectively, suggesting
that, in the asymptotic region of high feedback overhead, the number of
shape quantization bits should be approximately $(2M-1)$ times than the
number of gain quantization bits. The analysis and importance of bit
allocation is borne out by the simulation results that show significant
worse performance for the usual approach (in MU MIMO downlink systems)
of only quantizing the gain or shape but not both.

Our work with respect to the gain distortion calculation is quite
general, since the gain quantization distortion of other distributions
like Rician, Nakagami and Weibull fading can also be calculated using
Bennett's integral. However, the optimal bit allocation results might
be different from the Rayleigh fading case considered in this paper.


\appendices

\section{Proof of Lemma 1}


The authors of~\cite{Karasawa} have provided the following pdf of the
eigenvalues of a MIMO channel,
\begin{equation}
f(\lambda_e) = \frac{1}{\left(L(e)-1\right)!} \frac{\lambda_e^{L(e)-1}}{\beta^{L(e)}}
exp \left(-\frac{\lambda_e}{\beta} \right).    \label{eq:eig_dist1}
\end{equation}
Here, $\lambda_e$ denotes the $\mathrm{e}^{th}$ eigenvalue of the
Wishart matrix (i.e., $\mathbf{H}^H \mathbf{H}$ or $\mathbf{H}
\mathbf{H}^H$). $e$ denotes the index of the ordered eigenvalues. $L(e)
= (M - e) (N_k - e)$. $\beta$ is a constant whose value is given
through the following equation,
\begin{equation}
\beta = \frac{\tilde{\lambda_e}}{L(e)}.
\end{equation}
Here $\tilde{\lambda_e}$ is the mean of the eigenvalue.
\eqref{eq:eig_dist1} provides the probability distribution function of
the eigenvalue of the Wishart matrix, $\lambda_e$. In our proposed
algorithm, we are trying to quantize $g$, the singular values of the
Gaussian matrix $\mathbf{H}$. Now, $ \lambda_e = g^2 $.

Using Jacobian transformation~\cite{Strang}, the probability
distribution of the singular values of the Gaussian matrix can be found
as follows,
\begin{align}
f_g (r) & = \frac{1}{(L(e) - 1) !} \frac{(r^2)^{L(e)-1}}{\beta^{L(e)}} \exp
\left(-\frac{r^2}{\beta}\right) 2r. \label{eq:norm_dist1}   \allowdisplaybreaks[1]
\end{align}
Therefore,
\begin{align}
||f_g (r)||_{\frac{1}{3}} & = \frac{2}{(L(e) - 1) !} \frac{1}{\beta^{L(e)}} \left(\int_0^{\infty}
r^{\frac{2 L(e)-1}{3}} exp \left( - \frac{r^2}{3 \beta} \right) dr \right)^3 \allowdisplaybreaks[1].
\label{eq:norm_dist3}
\end{align}
Using standard mathematical tables of~\cite{Beyer} ( P - 380, eqn - 662), we find
\begin{equation}
\int_0^{\infty} x^n exp \left(-a x^p \right) dx = \frac{\Gamma \left(\frac{n+1}{p} \right)}
{p a^{\left(\frac{n+1}{p} \right)}}. \label{eq:beyer_int}
\end{equation}
Comparing \eqref{eq:beyer_int} with \eqref{eq:norm_dist3}, we find, $ n = \frac{2L(e) - 1}{3}$,
$a = \frac{1}{3 \beta}$, $p = 2$. Therefore,
\begin{align}
\left(\int_0^{\infty} r^{\frac{2 L(e)-1}{3}} exp \left( - \frac{r^2}{3 \beta} \right) dr \right) & =
\frac{\Gamma \left(\frac{\frac{2 L(e) - 1}{3} + 1}{2} \right)}
{2 \left(\frac{1}{3 \beta} \right)^{\frac{\frac{2 L(e) - 1}{3} + 1}{2}}}  \label{eq:beyer_int2}
\allowdisplaybreaks[1] \\
\left(\int_0^{\infty} r^{\frac{2 L(e)-1}{3}} exp \left( - \frac{r^2}{3 \beta} \right) dr \right) & =
\frac{1}{2} \left(3 \beta \right)^{\frac{L(e)+1}{3}} \Gamma \left(\frac{L(e)+1}{3}\right)
\label{eq:beyer_int3} \allowdisplaybreaks[1] \\
\left(\int_0^{\infty} r^{\frac{2 L(e)-1}{3}} exp \left( - \frac{r^2}{3 \beta} \right) dr \right)^3 & =
\frac{1}{8} \left(3 \beta \right)^{L(e)+1} \Gamma^3 \left(\frac{L(e)+1}{3}\right)
\allowdisplaybreaks[1].  \label{eq:beyer_int4}
\end{align}
Using \eqref{eq:beyer_int4} in \eqref{eq:norm_dist3}, we get,
\begin{align}
||f_g (r)||_{\frac{1}{3}} & = \frac{2}{(L(e) - 1)!} \frac{1}{\beta^{L(e)}} \frac{1}{8} 3^{L(e) + 1}
\beta^{L(e) + 1}
\Gamma^3 \left(\frac{L(e) + 1}{3} \right) \allowdisplaybreaks[1] \\
& = \frac{3 \times 3^{L(e)} \beta}{4 (L(e) - 1)!} \Gamma^3 \left(\frac{L(e) + 1}{3} \right).
\allowdisplaybreaks[1]
\end{align}
%



\section{Proof of Lemma 2}

Using~\eqref{eq:CCDF1_old},
\begin{align}
Pr [ \min_{i \in N} ||\mathbf{s}-\mathbf{\hat{s}}_i||^2 \geq b  ] & = \left(1 - \frac{(2M-1) C_{2M-1}
\int^{\cos^{-1} \left(1-0.5 b\right)}_0 \sin^{2M-2} \phi \, d\phi}{2M C_{2M}} \right)^N
\allowdisplaybreaks[1] \label{eq:CCDF1}  \\
& = \left(1 - K_1 \int^{\cos^{-1} \left(1-0.5 b\right)}_0 \sin^{2M-2} \phi \, d\phi \right)^N
\allowdisplaybreaks[1] \label{eq:CCDF2}  \\
& \approx \left(1 - K_1 \int^{\cos^{-1} \left(1-0.5 b\right)}_0 \phi^{2M-2} d\phi \right)^N
\allowdisplaybreaks[1]  \label{eq:CCDF3}  \\
& = \left(1 - K_2 \left({\cos^{-1} \left(1-0.5 b\right)}\right)^{2M-1} \right)^N.
\allowdisplaybreaks[1]  \label{eq:CCDF4}
\end{align}
In \eqref{eq:CCDF2}, we assumed $K_1 = \frac{(2M-1) C_{2M-1}}{2M
C_{2M}}$. \eqref{eq:CCDF3} follows from the fact that, given a large
number of quantization vectors, i.e., at high bit rate, the
complementary cumulative distribution function (CCDF) is significant
only for smaller values of $\phi$. For these smaller angles, we can
assume $ \sin \phi \approx \phi$. Equation \eqref{eq:CCDF4} follows
from assuming $K_2 = \frac{K_1}{2M-1}$.

\begin{figure}[t]
 \begin{center}
 \epsfig{figure=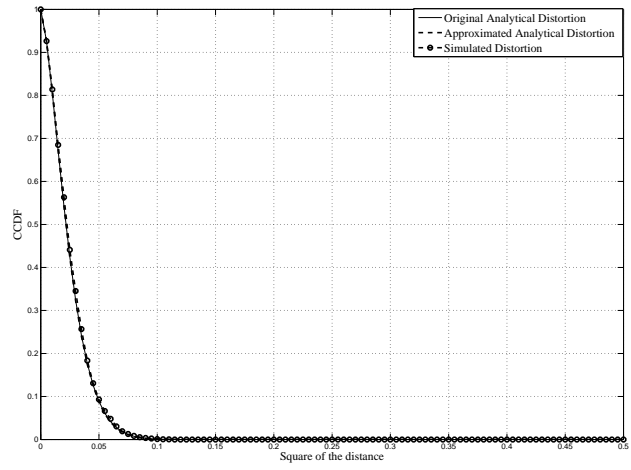,width=100mm}
 \end{center}
  \caption{Comparison of the original and approximated complementary cumulative distribution
  function of the shape distortion of a 2x1 vector (10 bit quantization)}
  \label{fig:shape_quantization_CCDF}
\end{figure}

Figure~\ref{fig:shape_quantization_CCDF} compares the simulated shape
distortion with the original and approximate analytical shape
distortion of a $2 \times 1$ $\mathbb{C}^{M}$ vector. Here, the CCDF of
the original and approximate analytical expressions are superimposed
with the simulated CCDF. Hence, \eqref{eq:CCDF2} and \eqref{eq:CCDF3}
accurately model the actual distortion. This justifies the transition
from \eqref{eq:CCDF2} to \eqref{eq:CCDF3}.

Now, using \eqref{eq:expect}, we find,
\begin{align}
E(b) & = \int_0^4 Pr [ \min_{i \in N} ||\mathbf{s}-\mathbf{\hat{s}}_i||^2 \geq b  ] db
\label{eq:expect1} \allowdisplaybreaks[1] \\
& = \int_0^a \left(1 - K_2 \left({\cos^{-1} \left(1-0.5 b\right)}\right)^{2M-1} \right)^N db
\label{eq:expect2}
\allowdisplaybreaks[1]  \\
& = 2 \int_0^{\psi} \left(1 - K_2 \theta^{2M-1} \right)^N \sin(\theta) d\theta \label{eq:expect3}
\allowdisplaybreaks[1]  \\
& \approx 2 \int_0^{\psi} \left(1 - K_2 \theta^{2M-1} \right)^N \theta d\theta \label{eq:expect4}
\allowdisplaybreaks[1] \\
& \approx 2 \int_0^{1} \left(1 - K_2 \theta^{2M-1} \right)^N \theta d\theta \label{eq:expect5}
\allowdisplaybreaks[1] \\
& = 2 \int_0^{1} \left( \sum^N_{i=0} {{N}\choose{i}} (-1)^i K_2^i \theta^{i(2M-1)+1} \right)
d\theta \label{eq:expect6} \allowdisplaybreaks[1] \\
& = 2 \sum^N_{i=0} \frac{{{N}\choose{i}} (-1)^i K_2^i}{i(2M-1)+2}. \label{eq:expect7}
\end{align}
The transition from \eqref{eq:expect1} to \eqref{eq:expect2} can be
explained as follows: the similarity between \eqref{eq:CCDF2} and
\eqref{eq:CCDF3} holds only for smaller values of $b$ since $ \sin \phi
\neq \phi $ for larger $ \phi $. Therefore, although the square of the
Euclidean distance between two random unit norm vectors can vary from 0
to 4, \eqref{eq:CCDF4} holds only for a smaller range of $b$. At the
presence of a large number of codewords, the squared distance between
the original and the quantized channel takes large values with a
negligibly small probability. Therefore, we can truncate the range of
$b$ as long as the CCDF of the original function is negligible outside
the range, i.e., the limited range of $b$ does not have any significant
affect on the calculation of the expected value of the distortion.
Using this analysis, in \eqref{eq:expect2}, we use $a$ as the truncated
range, i.e., we assume that $b$ can vary from $0$ to $a$.

In \eqref{eq:expect3} we assumed, $ \theta = \left({\cos^{-1}
\left(1-0.5 b\right)}\right)$. Therefore, $\psi = \left({\cos^{-1}
\left(1-0.5 a\right)}\right)$. Since only smaller angles of $\theta$
contribute to $E(b)$, we assumed $ \sin \theta \approx \theta$ in
\eqref{eq:expect4}. In \eqref{eq:expect6}, we assumed $ \psi = 1$ to
simplify the other calculations.

\begin{figure}[t]
 \begin{center}
 \epsfig{figure=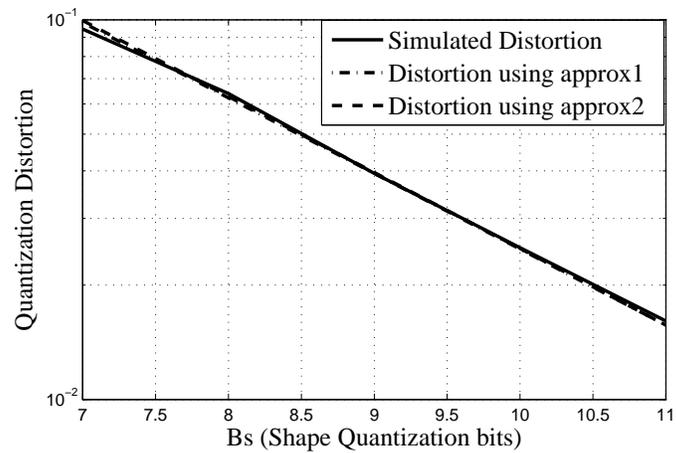,width=100mm}
 \end{center}
  \caption{Justification of the approximations used in Shape distortion calculation}
  \label{fig:Shape_CCDF_approx}
\end{figure}

Fig.~\ref{fig:Shape_CCDF_approx} justifies the approximations that we
used in the derivations of shape distortion calculation. Here, approx1
and approx2 denote $ \sin (\theta) \approx \theta$ (ref:
eq.~\ref{eq:expect4})  and $ \psi \approx 1$ (ref:
eq.~\ref{eq:expect5}) respectively. As Fig.~\ref{fig:Shape_CCDF_approx}
shows, the three curves are superimposed with each other. Therefore,
our justifications are valid for high bit rate quantization.

Applying $ {N \choose i} = \frac{(-1)^i (-N)_i}{i!} $, where $(-N)_i = \frac{\Gamma(-N+i)}
{\Gamma(-N)}$~\cite{Love:b}, \eqref{eq:expect7} takes the following form,
\begin{align}
\sum^N_{i=0} \frac{(-1)^i (-N)_i (-1)^i K_2^i}{i!(i(2M-1)+2)} & = \frac{2}{2M-1} \sum^N_{i=0}
\frac{(-N)_i K_2^i}{i! (i + \frac{2}{2M-1})}  \label{eq:pochman2}  \allowdisplaybreaks[1] \\
& = \frac{2}{2M-1} \frac{N!}{\frac{2}{2M-1} \left(1+\frac{2}{2M-1}\right)_N} K_2^{\frac{-2}{2M-1}}
\label{eq:pochman3}  \allowdisplaybreaks[1] \\
& = \frac{N! \Gamma \left(1+\frac{2}{2M-1} \right)}{\Gamma \left(N + 1 + \frac{2}{2M-1}\right)}K_3
\label{eq:pochman4} \allowdisplaybreaks[1] \\
& = \frac{N \Gamma(N) \Gamma\left(\frac{2M+1}{2M-1}\right)}{\Gamma \left(N + \frac{2M+1}{2M-1}\right)}
 K_3  \allowdisplaybreaks[1] \\
& = N \beta \left(N,\frac{2M+1}{2M-1} \right) K_3. \label{eq:pochman6} \allowdisplaybreaks[1]
\end{align}
\eqref{eq:pochman3} was found using (\cite{hansen}, 6.6.8). In
\eqref{eq:pochman4}, we assumed $K_3 = K_2^{-\frac{2}{2M-1}}$.
\eqref{eq:pochman6} was obtained using the relation between the gamma
and beta function, $ \beta (a,b) = \frac{\Gamma (a) \Gamma (b)}{\Gamma
(a+b)}$~\cite{Papoulis}. Following a similar work in~\cite{Jindal:a},
we find,
\begin{align}
N \beta \left(N,\frac{2M+1}{2M-1} \right) & = 2^B \frac{\Gamma(2^B) \Gamma(1+\frac{2}{2M-1})}
{\Gamma(2^B + 1+\frac{2}{2M-1})}  \label{eq:bound2}  \allowdisplaybreaks[1] \\
& \leq 2^B \frac{\Gamma(2^B)}{\Gamma(2^B + 1+\frac{2}{2M-1})}  \allowdisplaybreaks[1]
\label{eq:bound3}   \\
& = \frac{\Gamma(2^B+1)}{\Gamma(2^B + 1+\frac{2}{2M-1})}. \label{eq:bound4} \allowdisplaybreaks[1]
\end{align}
The preceding inequality in \eqref{eq:bound3} is justified by the
following reasoning: due to the convexity of the gamma
function~\cite{Jindal:a} and the fact that $ \Gamma (1) = \Gamma (2) =
1$, $ \Gamma(x) \leq 1$ for $ 1 \leq x \leq 2$ . Let, $y=2^B +
\frac{2}{2M-1}$, $ t = 1 - \frac{2}{2M-1}$, so that, $ y + t = 2^B +
1$, $y + 1 = 2^B + 1 + \frac{2}{2M-1}$. By applying Kershaw's
inequality for the gamma function~\cite{Kershaw},
\begin{equation}
\frac{\Gamma(y+t)}{\Gamma(y+1)} < \left(y+\frac{t}{2} \right)^{t-1} \,
            \forall \, y > 0 \, , \, 0 < t < 1.  \label{eq:Kershaw}
\end{equation}
Using \eqref{eq:Kershaw},
\begin{align}
\frac{\Gamma(2^B+1)}{\Gamma(2^B + 1+\frac{2}{2M-1})} & < \left(2^B + \frac{2}{2M-1} + 0.5 -
\frac{1}{2M-1} \right)^{\frac{-2}{2M-1}}  \label{eq:bound5} \allowdisplaybreaks[1] \\
& = \left(2^B + \frac{1}{2M-1} + 0.5\right)^{\frac{-2}{2M-1}}  \allowdisplaybreaks[1]
\label{eq:bound6} \\
& < 2^{\frac{-2B}{2M-1}}.    \label{eq:bound7} \allowdisplaybreaks[1]
\end{align}
Using \eqref{eq:bound7} and the value of $K_3$ we find,
\begin{equation}
2 \sum^N_{i=0} \frac{(-1)^i (-N)_i (-1)^i K_2^i}{i!(i(2M-1)+2)} <
\left(\frac{C_{2M-1}}{2MC_{2M}}\right)^{-\frac{2}{2M-1}} 2^{\frac{-2B}{2M-1}}.
\end{equation}
Using the values of $C_{2M-1}$ and $C_{2M}$ one can obtain,
\begin{equation}
E (b) < K_s 2^{\frac{-2 B_s}{2M-1}},   \label{eq:lemma_expectation_appendix}
\end{equation}
where,  $K_s = \left(\frac{\pi^{\frac{2M-1}{2}} \Gamma(M)}{2 \pi^M
\Gamma \left(\frac{2M-1}{2} + 1 \right)} \right)^ {\frac{-2}{2M-1}}$ is
a constant with respect to $B_s$.



\section{Proof of Theorem 1}

Taking the 1st and 2nd order derivatives of \eqref{eq:distortion}, we find,
\begin{align}
\frac{d D}{d B_s} & = \bar{K_s} (\ln 2) 2^{-\frac{2 B_s}{2 M - 1}} \left(-\frac{2}{2M-1} \right) +
K_g (\ln 2) \left( 2^{-2 (B - B_s)} \right) 2  \label{eq:1st_derivative} \allowdisplaybreaks[1] \\
\frac{d^2 D}{d^2 B_s} & = \bar{K_s} (\ln 2)^2  2^{-\frac{2 B_s}{2 M - 1}} \left(-\frac{2}{2M-1}
\right)^2 + K_g (2 \ln 2)^2 \left(2^{-2(B - B_s)} \right).  \label{eq:2nd_derivative}
\allowdisplaybreaks[1]
\end{align}
From \eqref{eq:2nd_derivative}, $ \frac{d^2 D}{d^2 B_s} \geq 0$.
Therefore, the optimal bit allocation problem is convex~\cite{Boyd}.
Now, equating the 1st derivative to be zero,
\begin{align}
\frac{\bar{K_s}}{2 M - 1} 2^{\frac{-2 B_s}{2 M - 1}} & =  K_g 2^{-2(B-B_s)} \allowdisplaybreaks[1] \\
2^{-2B + 2 B_s + \frac{2 B_s}{2M-1}} & = \frac{\bar{K_s}}{K_g (2M - 1)} \allowdisplaybreaks[1] \\
\frac{2 M B_s}{2 M - 1} & = B + \frac{1}{2} \log_2 \left(\frac{\bar{K_s}}{K_g (2M - 1)}\right)
\allowdisplaybreaks[1] \\
B_s & = \frac{2M-1}{2M} B + \frac{2M-1}{4M} \log_2 \left(\frac{\bar{K_s}}{K_g (2M - 1)}\right).
\allowdisplaybreaks[1]
\end{align}
Therefore, at the optimal point,
\begin{align}
B_s & = \frac{2M-1}{2M} B + \frac{2M-1}{4M} \log_2 \left(\frac{\bar{K_s}}{K_g (2M - 1)}\right)
\allowdisplaybreaks[1] \label{eq:shape_bits} \\
B_g & = \frac{1}{2M} B - \frac{2M-1}{4M} \log_2 \left(\frac{\bar{K_s}}{K_g (2M - 1)}\right).
\allowdisplaybreaks[1]  \label{eq:gain_bits}
\end{align}
%

\bibliographystyle{IEEEbib}
\bibliography{bib_letter}

\newpage
\begin{figure}[t]
 \begin{center}
 \epsfig{figure=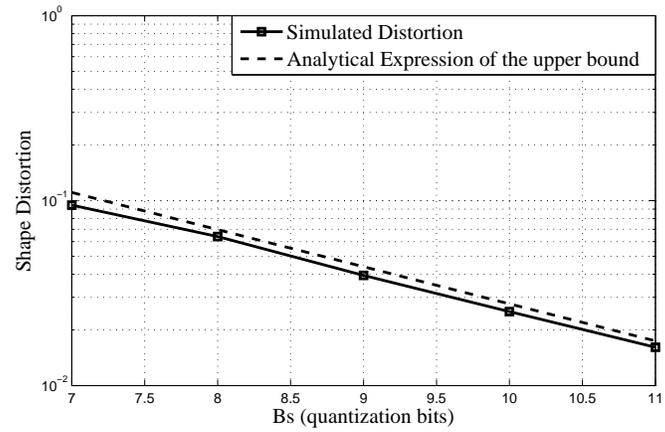,width=100mm}
 \end{center}
  \caption{Comparison of the simulated distortion with the theoretical upper bound
  (2x1 complex vector)}  \label{fig:shape_distortion_slope}
\end{figure} %

\end{document}